\documentclass[10pt,journal,epsfig]{IEEEtran}
\usepackage{graphicx}
\usepackage{subfigure,color}
\usepackage{amssymb}
\usepackage{cite,comment}
\usepackage{amsmath,setspace}
\usepackage{bm,comment,epstopdf}
\usepackage{algorithm,balance}
\usepackage{algpseudocode}
\makeatletter

\newcommand{\Rmnum}[1]{\expandafter\@slowromancap\romannumeral #1@}
\newcommand{\mv}[1]{\mbox{\boldmath{$ #1 $}}}

\makeatother

\newtheorem{remark}{\underline{Remark}}

\newtheorem{definition}{Definition}

\begin{document}
\title{\LARGE Intelligent Reflecting Surface for Multi-Path Beam Routing with Active/Passive Beam Splitting and Combining}
\author{Weidong Mei, \IEEEmembership{Member, IEEE}, and Rui Zhang, \IEEEmembership{Fellow, IEEE}
\thanks{The authors are with the Department of Electrical and Computer Engineering, National University of Singapore, Singapore 117583 (e-mails: \{wmei, elezhang\}@nus.edu.sg).}}
\maketitle

\begin{abstract}
Intelligent reflecting surface (IRS) can be densely deployed in wireless networks to significantly enhance the communication channels. In this letter, we consider the downlink transmission from a multi-antenna base station (BS) to a single-antenna user, by exploiting the cooperative passive beamforming (CPB) and line-of-sight (LoS) path diversity gains of multi-IRS signal reflection. Unlike existing works where only one single multi-IRS reflection path from the BS to user is selected, we propose a new and more general {\it \textbf{multi-path beam routing}} scheme. Specifically, the BS sends the user's information signal via multiple orthogonal active beams (termed as {\it \textbf{active beam splitting}}), which point towards different IRSs. Then, these beamed signals are subsequently reflected by selected IRSs via their CPB in different paths, and finally coherently combined at the user's receiver (thus named {\it \textbf{passive beam combining}}). For this scheme, we formulate a new multi-path beam routing design problem to jointly optimize the number of IRS reflection paths, the selected IRSs for each of the reflection paths, the active/passive beamforming at the BS/each selected IRS, as well as the BS's power allocation over different active beams, so as to maximize the received signal power at the user. To solve this challenging problem, we first derive the optimal BS/IRS beamforming and BS power allocation for a given set of reflection paths. The clique-based approach in graph theory is then applied to solve the remaining multi-path selection problem efficiently. Simulation results show that our proposed multi-path beam routing scheme significantly outperforms its conventional single-path beam routing special case.
\end{abstract}
\begin{IEEEkeywords}
	Intelligent reflecting surface, multi-path beam routing, beam splitting, beam combining, graph theory.
\end{IEEEkeywords}

\section{Introduction}
In comparison with today's fifth-generation (5G) wireless network, the sixth-generation (6G) wireless network in the future is expected to pose more stringent requirement on the network's key performance indicators (KPIs), such as data rate, reliability, coverage, energy efficiency, etc. Recently, intelligent reflecting surface (IRS) has emerged as a promising technique to boost the wirelesses network performance, so as to meet the demanding KPIs of 6G cost-effectively\cite{wu2020intelligent}. Specifically, by jointly tuning the phase shifts of a large number of reflecting elements, IRS is able to reshape the wireless channels flexibly for communication performance enhancement. As IRS only reflects the impinging signal as a passive array, it incurs much lower hardware cost and energy consumption as compared to conventional active transceivers/relays.

The appealing advantages of IRS have spurred a great deal of interest in investigating its performance gains under different wireless system setups (see e.g., \cite{wu2020intelligent,wu2019towards,di2019smart} and the references therein). However, existing works on IRS mainly considered the single-reflection links between the base station (BS) and users aided by one or multiple IRSs, but ignored the multi-reflection links over different IRSs, which can be utilized to further refine the end-to-end wireless channel between the BS and each user. For example, by deploying two IRSs at the sides of the BS and user, respectively, and exploiting the inter-IRS line-of-sight (LoS) channel, it was shown in \cite{han2020cooperative} that a higher cooperative passive beamforming (CPB) gain can be achieved, which increases {\it quartically} with the total number of IRS reflecting elements, as compared to the {\it quadratic} passive beamforming gain by the conventional single-IRS reflection link\cite{wu2020intelligent}. Motivated by this, some other works\cite{zheng2021double,dong2021double,you2021wireless,zheng2021efficient} have recently investigated the optimal CPB design and efficient channel estimation for the double-IRS aided systems. 

\begin{figure}[!t]
\centering
\includegraphics[width=3.5in]{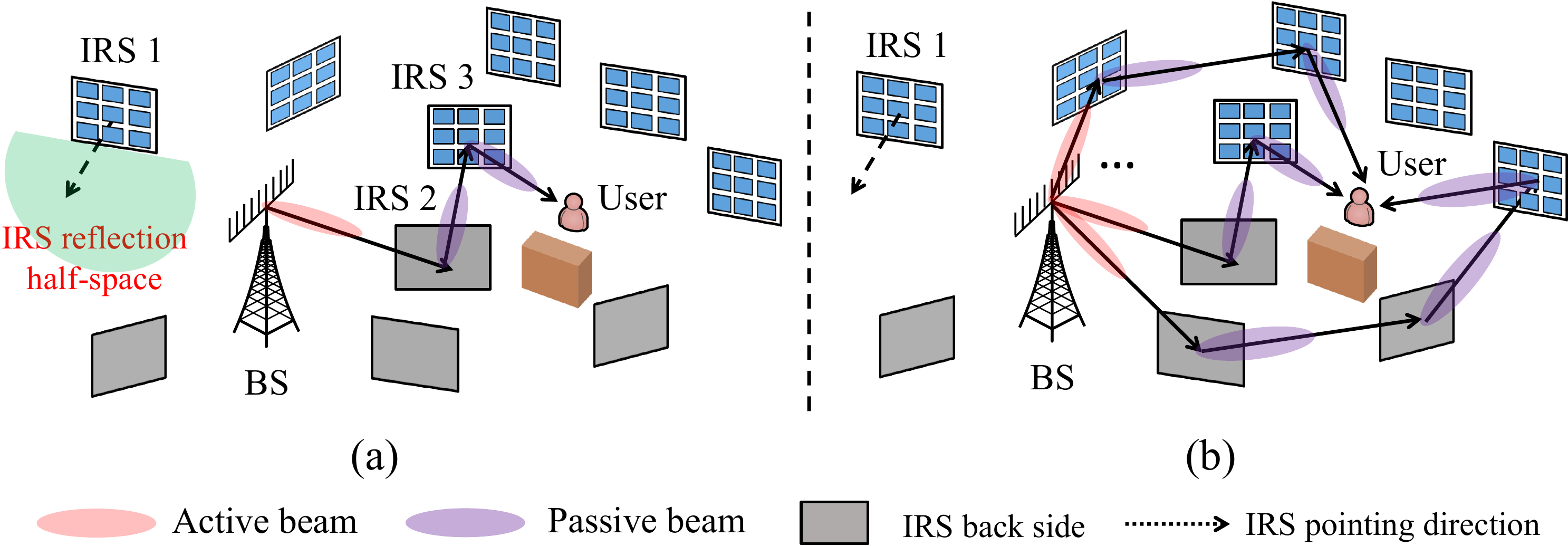}
\DeclareGraphicsExtensions.
\vspace{-12pt}
\caption{(a) Conventional single-path beam routing scheme, and (b) Proposed multi-path beam routing scheme with active/passive beam splitting/combining.}\label{split}
\vspace{-15pt}
\end{figure}
More generally, with multiple (more than two) IRSs engaged to assist a wireless link, their multi-reflection LoS-dominant channel can provide an even more pronounced CPB gain which well compensates the higher multiplicative path loss due to the increased number of IRS reflections. Besides, in a complex environment with scattered obstacles, more IRSs can offer a higher path diversity gain by establishing multiple blockage-free LoS links between the BS and each blocked user. This thus leads to a new passive beam routing design problem, with a goal to select the optimal BS-user reflection path and optimize the active/passive beamforming at the BS/IRSs in the selected path, such that the end-to-end BS-user channel power gain is maximized. By assuming perfect channel knowledge on all links, the authors in \cite{huang2021multi,liang2021cascaded} studied the optimal active/passive beamforming design at the BS/IRSs for a fixed BS-user reflection path, while the joint path selection and beamforming design problem was investigated in \cite{mei2020cooperative} and \cite{mei2021mbmh} in the cases of single and multiple users, respectively. Furthermore, the authors in \cite{mei2021distributed} proposed a distributed beam training scheme to realize such joint design in practice, which dispenses with sophisticated channel estimation. However, in the above works, only a single reflection path is selected for each user (see, e.g., the reflection path via IRSs 2 and 3 in Fig.\,\ref{split}(a)), while all other IRSs that are not in the selected path are simply switched off. It is evident that this approach may not fully exploit all the available IRSs and their path diversity for enhancing the user performance.

To tackle this problem, we propose in this letter a new and more general {\it multi-path beam routing} scheme. Specifically, as shown in Fig.\,\ref{split}(b), the multi-antenna BS transmits the user's message via multiple orthogonal active beams pointing towards different IRSs (named {\it active beam splitting}). Then, these beamed signals are successively reflected by a set of selected IRSs in different paths and finally coherently combined at the user's receiver (named {\it passive beam combining}). As such, more IRSs and their CPB as well as path diversity gains can be exploited in the proposed multi-path beam routing scheme, as compared to its special case of single-path beam routing \cite{mei2020cooperative,mei2021mbmh,mei2021distributed} shown in Fig.\,\ref{split}(a). For this new scheme, we aim to jointly optimize the number of reflection paths, selected IRSs for each path, active/passive beamforming at the BS/each IRS, and the BS's power allocation over the active beams to maximize the user's received signal power. To solve this challenging problem, we first derive the optimal BS/IRS active/passive beamforming and the BS's power allocation for a given set of reflection paths, by exploiting the inter-IRS LoS channels. Then, the clique-based approach in graph theory is applied to solve the remaining multi-path selection problem efficiently. Numerical results demonstrate that our proposed scheme achieves significant performance improvement over its conventional single-path counterpart.\vspace{-9pt}

\section{System Model}
As shown in Fig.\,\ref{split}, we consider the downlink transmission in a multi-IRS aided system in a given time slot, where a BS equipped with $N_B$ antennas communicates with a single-antenna user scheduled in this time slot with the help of $J$ distributed IRSs in the set ${\cal J}\triangleq \{1,2,\cdots,J\}$, each equipped with $M$ reflecting elements. Due to the dense obstacles in the environment, the BS can only communicate with the user via one or more IRS reflection paths between them, each formed by a subset of IRSs which successively reflect the signal from the BS to the user. To ensure the path diversity, the number of IRSs should be sufficiently large and they should be densely deployed in the region of interest. Let ${\mv w}_B \in {\mathbb C}^{N_B \times 1}$ denote the active beamforming vector employed by the BS. To ease practical implementation, we consider that the BS employs a discrete beamforming codebook comprising orthogonal and unit-power beams, denoted as ${\cal W}_B$, i.e., ${\mv w}_B \in {\cal W}_B$. Let ${\mv \Phi}_j={\rm diag}({\mv \theta}_j) \in {\mathbb C}^{M \times M}$ denote the reflection matrix of IRS $j, j \in \cal J$, where ${\mv \theta}_j \in {\mathbb C}^{M \times 1}$ denotes its passive beamforming vector and is assumed to be selected from a passive beamforming codebook ${\cal W}_I$, i.e., ${\mv \theta}_j \in {\cal W}_I, j \in \cal J$. For convenience, we refer to the BS and user as nodes 0 and $J+1$ in the system, respectively. Moreover, we assume that the BS and each IRS are equipped with a uniform linear and rectangular array, respectively. The distance between any two nodes $i$ and $j$ in the system is denoted as $d_{i,j}$, given a reference element selected at the BS/each IRS.

Since each IRS can only achieve 180$^\circ$ half-space reflection, we define a pointing direction for each IRS which is perpendicular to its reflecting surface and points to its reflection half-space, as shown in Fig.\,\ref{split}. Thus, to effectively reflect the signal from the BS to the user in any reflection path, both the previous and next nodes of each intermediate IRS should be located in its reflection half-space. Moreover, we only consider the reflection paths where the BS's signal is reflected from one IRS to a farther IRS from the BS (in the downlink) but not the other way around, as the latter in general results in more severe path loss. For example, in Fig.\,\ref{split}, the path successively over IRSs 3 and 2 leads to much higher ``product'' path loss than that successively over IRSs 2 and 3. 

To enhance the strength of BS-user reflection links, we consider that only the LoS links in the system are exploited for IRS signal reflection, while treating all non-LoS (NLoS) links as part of environment scattering, which generally has a marginal effect on the user performance\cite{mei2021mbmh,mei2021distributed}. Accordingly, to depict all LoS paths from the BS to the user, we define an LoS graph for all nodes and their wireless links in the system, denoted as $G_L=(V_L,E_L)$, where $V_L=\{0,1,2,\cdots,J+1\}$ and $E_L$ denote the sets of vertices and edges in $G_L$, respectively. In particular, there exists an edge from vertex $i$ to vertex $j$, denoted as $e_{i,j}$, if there is an LoS path between nodes $i$ and $j$ and the following two conditions are satisfied: 1) effective signal reflection can be achieved from node $i$ to node $j$ subject to the half-space constraint at each IRS; and 2) node $j$ is farther away from the BS than node $i$, i.e., $d_{0,j} > d_{0,i}$, to achieve outward signal reflection (except that node $j$ is the user). It follows that each LoS path from the BS to the user corresponds to a path from vertex 0 to vertex $J+1$ in $G_L$. In this letter, we assume that the LoS graph $G_L$ is available at the BS by separate beam training\cite{mei2021distributed}.

Let ${\mv H}_{0,j} \in {\mathbb C}^{M \times N}, j \in {\cal J}$ denote the channel from the BS to IRS $j$, ${\mv S}_{i,j} \in {\mathbb C}^{M \times M}, i,j \in {\cal J}, i \ne j$ denote that from IRS $i$ to IRS $j$, and ${\mv g}_{j,J+1}^{H} \in {\mathbb C}^{1 \times M}, j \in {\cal J}$ denote that from IRS $j$ to the user. If there exists an LoS-dominant path from the BS to IRS $i$, i.e., $e_{0,i} \in E_L$, their channel can be approximated as ${\mv H}_{0,j} = {\sqrt \beta}d^{-1}_{0,j}{\tilde{\mv h}}_{j,2}{\tilde{\mv h}}^H_{j,1}$, where ${\tilde{\mv h}}_{j,1}$ and ${\tilde{\mv h}}_{j,2}$ denote the array responses at the BS and IRS $j$, respectively, and $\beta$ denotes the reference path gain at the reference distance of 1 meter (m). Similarly, if $e_{i,j} \in E_L, i,j \in \cal J$ and $e_{j,J+1} \in E_L, i \in \cal J$, we have ${\mv S}_{i,j} = {\sqrt \beta}{d^{-1}_{i,j}}{\tilde{\mv s}}_{i,j,2}{\tilde{\mv s}}^H_{i,j,1}$ and ${\mv g}^H_{j,J+1} = {\sqrt \beta}{d^{-1}_{j,J+1}}{\tilde{\mv g}}^H_{j,J+1}$, respectively, where ${\tilde{\mv s}}_{i,j,1}$, ${\tilde{\mv s}}_{i,j,2}$, and ${\tilde{\mv g}}^H_{i,J+1}$ are their corresponding array responses. Note that the array responses at each IRS can be further decomposed as the product of two component array responses in the horizontal and vertical dimensions, respectively, according to \cite{mei2020cooperative,mei2021mbmh}, for which the details are omitted for brevity. 

Based on the above, we can characterize the end-to-end channel from the BS to the user in any given multi-reflection LoS path, denoted as $\Omega=\{a_1,a_2,\cdots,a_L\}$, where $L \ge 1$ and $a_l \in \cal J$ denote the number of IRSs in $\Omega$ and the index of the $l$-th IRS, respectively. Then, the BS-user end-to-end channel under a given $\Omega$ is expressed as
\begingroup
\allowdisplaybreaks
\begingroup\makeatletter\def\f@size{9.3}\check@mathfonts
\def\maketag@@@#1{\hbox{\m@th\normalsize\normalfont#1}}%
\begin{equation}\label{recvsig}
h_{0,J+1}(\Omega)\!=\!{\mv g}^H_{a_L,J+1}{\mv \Phi}_{a_L}\Big(\prod\limits_{l=1}^{L-1}{\mv S}_{a_l,a_{l+1}}{\mv \Phi}_{a_l}\Big){\mv H}_{0,a_1}{\mv w}_B.
\end{equation}\normalsize

In our previous works \cite{mei2020cooperative,mei2021mbmh,mei2021distributed}, we have shown how to select the optimal reflection path $\Omega$ (as well as the corresponding active/passive beamforming design at the BS/IRSs in $\Omega$) among all feasible paths to maximize the end-to-end channel power gain, $\lvert h_{0,J+1}(\Omega) \rvert^2$. However, as shown in Fig.\,\ref{split}(a), this single-path beam routing solution may fail to exploit other IRSs which are not included in $\Omega$, while they can be utilized to further enhance the user's performance. To improve over this single-path beam routing scheme, we propose a new multi-path beam routing scheme, as detailed next.\vspace{-3pt}

\section{Multi-Path Beam Routing with Active/Passive Beam Splitting/Combining}
In the proposed multi-path beam routing scheme, multiple reflection paths are utilized to assist in the BS-user communication at the same time, as shown in Fig.\,\ref{split}(b). Specifically, the multi-antenna BS splits its precoding vector, ${\mv w}_B$, for transmitting the user's information signal, into multiple orthogonal active beams (i.e., active beam splitting), which are selected from its beamforming codebook ${\cal W}_B$ with different power allocations and point to different IRSs. Then, these beamed signals are successively reflected by selected IRSs in different paths, respectively. At last, the signals from different paths are coherently combined at the user's receiver (i.e., passive beam combining). The details are presented as follows.\vspace{-6pt}

\subsection{Active Beam Splitting}
Let $Q \;(\ge 1)$ and ${\mv w}_{B,q} \in {\cal W}_I$ denote the number of active orthogonal beams at the BS (or equivalently, the number of reflection paths) and the $q$-th beamforming vector, with $\lVert {\mv w}_{B,q} \rVert=1, q \in {\cal Q}=\{1,2,\cdots,Q\}$. Then, assuming unit transmit power at the BS, its precoding vector is expressed as
\begin{equation}\label{newabf1}
{\mv w}_B={\sum\nolimits_{q=1}^Q\sqrt{\alpha_q}{\mv w}_{B,q}},
\end{equation}
where $\alpha_q$ denotes the fraction of total transmit power allocated to the $q$-th orthogonal active beam with $\sum\nolimits_{q=1}^Q \alpha_q=1$. Assume that the $q$-th beam is associated with (reflected by) the IRSs in the $q$-th path, denoted as $\Omega_q=\{a^{(q)}_1,a^{(q)}_2,\cdots,a^{(q)}_{L_q}\}, q \in {\cal Q}$, where $a^{(q)}_l$ and $L_q$ denote the index of the $l$-th IRS and the number of IRSs in the $q$-th path, respectively. To ensure that each constituent link of $\Omega_q$ consists of an LoS path, it must hold that $e(a^{(q)}_l,a^{(q)}_{l+1}) \in E_L, \forall l=0,1,\cdots,L_q, q \in {\cal Q}$, where we assume $a^{(q)}_0=0$ and $a^{(q)}_{L_q+1}=J+1$, corresponding to the BS and user, respectively. Moreover, to simplify the IRS reflection design and reap a full CPB gain in each path, we consider that all paths $\Omega_q, q \in \cal Q$ have no common nodes with each other (except for the BS and user), i.e., $\Omega_q \cap \Omega_{q'} =\emptyset, q,q' \in {\cal Q}, q \ne q'$.

Similar to (\ref{recvsig}), for each active beam ${\mv w}_{B,q}$, its end-to-end channel with the user in its associated reflection path $\Omega_q, q \in \cal Q$ is given by
\begingroup\makeatletter\def\f@size{8}\check@mathfonts
\def\maketag@@@#1{\hbox{\m@th\normalsize\normalfont#1}}%
\begin{align}
h_{0,J+1}(\Omega_q)&\!=\!{\mv g}^H_{a^{(q)}_{L_q},J+1}{\mv \Phi}_{a^{(q)}_{L_q}}\Bigg(\prod\limits_{l=1}^{L_q-1}{\mv S}_{a^{(q)}_l,a^{(q)}_{l+1}}{\mv \Phi}_{a^{(q)}_l}\!\!\Bigg){\mv H}_{0,a^{(q)}_1}{\mv w}_{B,q} \nonumber\\
&=(\sqrt\beta)^{L_q+1}\prod\limits_{l=0}^{L_q}d^{-1}_{a^{(q)}_l,a^{(q)}_{l+1}}\Big(\prod\limits_{l=1}^{L_q}A^{(q)}_l\Big){\tilde{\mv h}}^H_{a^{(q)}_1,1}{\mv w}_{B,q}\label{recvsig2}
\end{align}\endgroup
where
\begin{equation}\label{Al}
A^{(q)}_l = \begin{cases}
	{\tilde{\mv s}}^H_{a^{(q)}_1,a^{(q)}_2,1}{\mv \Phi}_{a^{(q)}_1}{\tilde{\mv h}}_{a^{(q)}_1,2} &{\text{if}}\;\;l=1\\
	\tilde{\mv g}^H_{a^{(q)}_{L_q},J+1}{\mv \Phi}_{a^{(q)}_{L_q}}{\tilde{\mv s}}_{a^{(q)}_{L_q-1},a^{(q)}_{L_q},2} &{\text{if}}\;\;l=L_q\\
	{\tilde{\mv s}}^H_{a^{(q)}_l,a^{(q)}_{l+1},1}{\mv \Phi}_{a^{(q)}_l}{\tilde{\mv s}}_{a^{(q)}_{l-1},a^{(q)}_l,2} &{\text{otherwise}}.
	\end{cases}
\end{equation}

To maximize the BS-user effective channel power gain, i.e., $\lvert h_{0,J+1}(\Omega_q) \rvert^2$, it can be shown that the optimal ${\mv w}_{B,q}$ and passive beamforming at each IRS $a^{(q)}_l$ in $\Omega_q$ are given by\cite{mei2021mbmh}
\begingroup\makeatletter\def\f@size{9.2}\check@mathfonts
\def\maketag@@@#1{\hbox{\m@th\normalsize\normalfont#1}}%
\begin{align}
	{\mv w}^{\star}_{B,q} &= \arg \mathop{\max}\limits_{{\mv w} \in {\cal W}_B} \lvert{\tilde{\mv h}}^H_{a^{(q)}_1,1}\mv w \rvert, \label{abf}\\
	{\mv \theta}^{\star}_{a^{(q)}_l}&\!=\!
\begin{cases}
	\arg \mathop{\max}\limits_{{\mv \theta} \in {\cal W}_I}\lvert{\tilde{\mv s}}^H_{a^{(q)}_1,a^{(q)}_2,1}{\rm diag}(\mv \theta){\tilde{\mv h}}_{a^{(q)}_1,2}\rvert &{\text{if}}\;l=1\\
	\arg \mathop{\max}\limits_{{\mv \theta} \in {\cal W}_I}\lvert\tilde{\mv g}^H_{a^{(q)}_{L_q},J+1}{\rm diag}(\mv \theta){\tilde{\mv s}}_{a^{(q)}_{L_q-1},a^{(q)}_{L_q},2}\rvert \!\!&{\text{if}}\;l=L_q\\
	\arg \mathop{\max}\limits_{{\mv \theta} \in {\cal W}_I}\lvert{\tilde{\mv s}}^H_{a^{(q)}_l,a^{(q)}_{l+1},1}{\rm diag}(\mv \theta){\tilde{\mv s}}_{a^{(q)}_{l-1},a^{(q)}_l,2}\rvert \!\!&{\text{otherwise}},\label{pbf}
\end{cases}
\end{align}\endgroup
respectively. In practice, the optimal beamforming design in (\ref{abf}) and (\ref{pbf}) can be obtained by exploiting the cooperative beam training among the nodes without any explicit channel estimation\cite{mei2021distributed}, and the training among the BS and IRSs can be conducted offline thanks to the (nearly) constant BS-IRS and inter-IRS channels. It is also worth noting that although the codebook-based IRS passive beamforming has been studied in prior works (e.g., \cite{jamali2021power,najafi2021physics}), they considered the single-IRS system only.
\normalsize
\begin{remark}
The beamforming design in (\ref{abf}) and (\ref{pbf}) may result in inter-beam interference via the LoS links between any two reflection paths due to the side-lobe of each active or passive beam, which may affect the channel power gains, $\lvert h_{0,J+1}(\Omega_q) \rvert^2, q \in \cal Q$, since such ``leaked'' information signals are generally uncontrolled and may add to their controlled counterparts in a constructive or destructive manner at the receiver. However, as will be shown in Section \ref{sim}, the strength of such inter-beam interference is negligible if the number of antennas/elements at the BS/IRSs and the size of their codebooks are sufficiently large\cite{ngo2014aspects,perovic2020channel}, as this increases their angular resolution and thus reduces the side-lobe effects.  
\end{remark}
\vspace{-9pt}

\subsection{Passive Beam Routing and Combining}
Let $\bar A_l^{(q)}$ denote the value of $A_l^{(q)}$ in (\ref{Al}) with the optimal passive beamforming design in (\ref{pbf}). Then, by substituting (\ref{abf}) and (\ref{pbf}) into (\ref{recvsig2}), the end-to-end channel in (\ref{recvsig2}) becomes
\[
{\bar h}_{0,J+1}(\Omega_q)\!=\!(\sqrt\beta)^{L_q+1}\prod\limits_{l=0}^{L_q}d^{-1}_{a^{(q)}_l,a^{(q)}_{l+1}}\Big(\prod\limits_{l=1}^{L_q}{\bar A}^{(q)}_l\Big){\tilde{\mv h}}^H_{a^{(q)}_1,1}{\mv w}^{\star}_{B,q}.
\]
Due to the negligible inter-beam/-path interference, the received information signal at the user's receiver in each path $\Omega_q$ is approximately equal to $y_q=\sqrt{\alpha_q}{\bar h}_{0,J+1}(\Omega_q)x_B, q \in \cal Q$, where $x_B$ denotes the transmitted symbol by the BS and we assume $x_B=1$ for notational brevity.

However, the received signals over the $Q$ paths, i.e., $y_q, q \in \cal Q$, may not coherently add at the user's receiver due to their random channel phases as a result of different delays. To maximize the received signal power at the user, they should be coherently combined at the user's receiver. To this end, the BS can append a phase rotation, $e^{-j\angle({\bar h}_{0,J+1}(\Omega_q))}$, to each active beam ${\mv w}^{\star}_{B,q}$ to compensate for the end-to-end delay in $\Omega_q$. The phase/delay information, $\angle({\bar h}_{0,J+1}(\Omega_q))$, can be computed by the BS based on the distributed beam training among the nodes and their feedback information\cite{mei2021distributed}. As such, the received signal (excluding the noise) power at the user is given by
\begin{equation}\label{eq1}
\Gamma_u = \left(\sum\limits_{q=1}^Q\lvert y_q \rvert\right)^2=\left(\sum\limits_{q=1}^Q\sqrt{\alpha_q}\lvert \bar h_{0,J+1}(\Omega_q) \rvert\right)^2.
\end{equation}

It is observed from (\ref{eq1}) that $\Gamma_u$ depends on the BS's power allocations over the $Q$ active beams. By applying the Cauchy–Schwarz inequality to (\ref{eq1}), we have
\begingroup\makeatletter\def\f@size{9.2}\check@mathfonts
\def\maketag@@@#1{\hbox{\m@th\normalsize\normalfont#1}}%
\begin{equation}\label{eq2}
\Gamma_u \le \left(\sum\limits_{q=1}^Q {\alpha_q} \right)\left(\sum\limits_{q=1}^Q\lvert {\bar h}_{0,J+1}(\Omega_q) \rvert^2\right)=\sum\limits_{q=1}^Q\lvert {\bar h}_{0,J+1}(\Omega_q) \rvert^2,
\end{equation}
where the optimal power allocations that achieve (\ref{eq2}) are given by $\alpha^{\star}_q=\lvert {\bar h}_{0,J+1}(\Omega_q) \rvert^2/\sum\nolimits_{q=1}^Q\lvert {\bar h}_{0,J+1}(\Omega_q) \rvert^2, q \in \cal Q$. It is noted that the maximum received signal power in (\ref{eq2}) is equal to the sum of the effective channel power gains in the $Q$ reflection paths, $\Omega_q, q \in {\cal Q}$, thanks to the coherent signal combining at the user.
\normalsize
\begin{remark}
It is worth mentioning that in addition to active beam splitting at the BS, {\it passive} beam splitting can also be realized at each IRS, by e.g., dividing it into multiple subsurfaces that reflect the incident signals to different nearby IRSs in different directions, thereby creating more split signal paths. However, we argue that passive beam splitting may not be efficient in practice due to the following reasons. First, the inter-beam interference may have a larger effect on the user performance due to the smaller size of each subsurface, which has a more severe side-lobe effect. Second, the smaller size of each subsurface also significantly reduces the CPB gain achievable in each reflection path, which may not compensate for the high path loss, even with the coherent combining of all split signal paths at the user. Third, the cooperative beam training needs to be performed for each subsurface, thus greatly increasing the codebook size at each IRS and thus the overall training complexity\cite{jamali2021power}.\vspace{-6pt}
\end{remark}

\section{Problem Formulation and Solution}
In this section, we formulate the multi-path beam routing optimization problem for the proposed scheme and propose an efficient solution to solve it.\vspace{-6pt}

\subsection{Problem Formulation}
In this letter, we aim to maximize the received signal power $\Gamma_u$ in (\ref{eq2}), by jointly optimizing the number of reflection paths $Q$ and selected IRSs in each path $\{\Omega_q\}_{q \in \cal Q}$, subject to the constraints that there is no common IRS between any two different paths. The optimization problem is formulated as
\begin{align}\label{op1}
{\text{(P1)}} \mathop {\max}\limits_{Q,\{\Omega_q\}_{q \in \cal Q}}\;&\sum\limits_{q=1}^Q\lvert {\bar h}_{0,J+1}(\Omega_q) \rvert^2 \nonumber\\
\text{s.t.}\;\;&e(a^{(q)}_l,a^{(q)}_{l+1}) \in E_L, \forall l=0,1,\cdots,L_q, q \in {\cal Q},\nonumber\\
&\Omega_q \cap \Omega_{q'} =\emptyset, q,q' \in {\cal Q}, q \ne q'.
\end{align}
Note that in practice, (P1) can be solved at the BS based on distributed IRS training and feedback information\cite{mei2021distributed}.

It is also noted that by setting $Q=1$ in (P1), it reduces to the conventional single-path beam routing problem \cite{mei2020cooperative,mei2021mbmh,mei2021distributed}, as shown in Fig.\,\ref{split}(a). As such, the optimal value of (P1) should be no smaller than that of the single-path beam routing special case. However, (P1) is a more challenging combinatorial optimization problem to solve, since the selection of different paths is coupled even with a fixed $Q$. To tackle this difficult problem, we propose an efficient graph-based solution for it in the next.\vspace{-6pt}

\subsection{Proposed Solution}
First, we present the following definition to transform (P1) into an equivalent graph-optimization problem.
\begin{definition}\label{vd}
In a graph, two paths are {\it vertex-disjoint} if they share no vertices in common except the start and end vertices.	
\end{definition}

Based on Definition \ref{vd}, since each reflection path $\Omega_q, q \in \cal Q$ corresponds to one path from vertex 0 to vertex $J+1$ in the LoS graph $G_L$, each feasible solution to (P1) corresponds to a set of vertex-disjoint paths from vertex 0 to vertex $J+1$ in $G_L$. However, optimization of disjoint paths generally results in an NP-complete problem\cite{mei2021mbmh}. 

To circumvent such difficulty, we apply a similar clique-based approach (CBA) as in \cite{mei2021mbmh} to solve (P1) via partial enumeration, for which only the main ideas are outlined below due to the space limit. {\it First, among all paths from vertex 0 to vertex $J+1$ in $G_L$, we select $K\; (\ge 1)$ candidate paths which achieve the $K$ largest end-to-end channel power gains by applying the Yen's algorithm\cite{west1996introduction}}. Denote by ${\cal K}$ the set of all candidate paths. It is evident that the optimal path in the single-path beam routing special case (corresponding to $K=1$), denoted as $\Omega_{\text{single}}^{\star}$, must be included in ${\cal K}$. From the paths in $\cal K$, we aim to find the best set of vertex-disjoint paths that achieves the maximum objective value of (P1). It is not difficult to see that if there exists at least one path in ${\cal K}$ which is vertex-disjoint from $\Omega_{\text{single}}^{\star}$ (e.g., $\Omega'$ with $\Omega' \cap \Omega_{\text{single}}^{\star}=\emptyset$), then the proposed multi-path beam routing scheme should outperform its single-path special case, as $\Omega'$ and $\Omega_{\text{single}}^{\star}$ can be employed to serve the user at the same time and thereby improve its performance. Nonetheless, the best set of vertex-disjoint paths may not necessarily include $\Omega_{\text{single}}^{\star}$, as will be shown in Section \ref{sim}. While if there is no path in ${\cal K}$ which is vertex-disjoint from $\Omega_{\text{single}}^{\star}$ (e.g., when the BS/user can only achieve an LoS link with one IRS), then the proposed multi-path beam routing scheme is not ensured to outperform its single-path special case. 

{\it Second, to obtain the best set of vertex-disjoint paths, we can construct a path graph $G_p$ for the candidate paths in $\cal K$ and then apply the Bron-Kerbosch algorithm\cite{west1996introduction} to enumerate the maximal cliques in $G_p$, each corresponding to a feasible solution to (P1)}. The required worst-case computational complexity is in the order of $3^{K/3}$. It follows that if the number of candidate paths $K$ is set to be sufficiently large, such that the optimal solution to (P1) is included in $\cal K$, then the proposed approach is able to achieve an optimal performance, at the cost of a higher time complexity. Nonetheless, as will be shown in Section \ref{sim}, it generally only requires a small $K$ to achieve the optimal performance for a moderate number of IRSs. {\it Finally, we compare the objective value of (P1) achieved by each maximal clique in $G_p$ and select the best one as the output.} It is worth mentioning that although the CBA is also applied in \cite{mei2021mbmh} for multi-path selection, it studies a special case of ours where only a single path is selected for each user without the active/passive beam splitting and combining designs.\vspace{-9pt}

\section{Simulation Results}\label{sim}
\begin{figure}[!t]
\centering
\subfigure[3D plot.]{\includegraphics[width=0.24\textwidth]{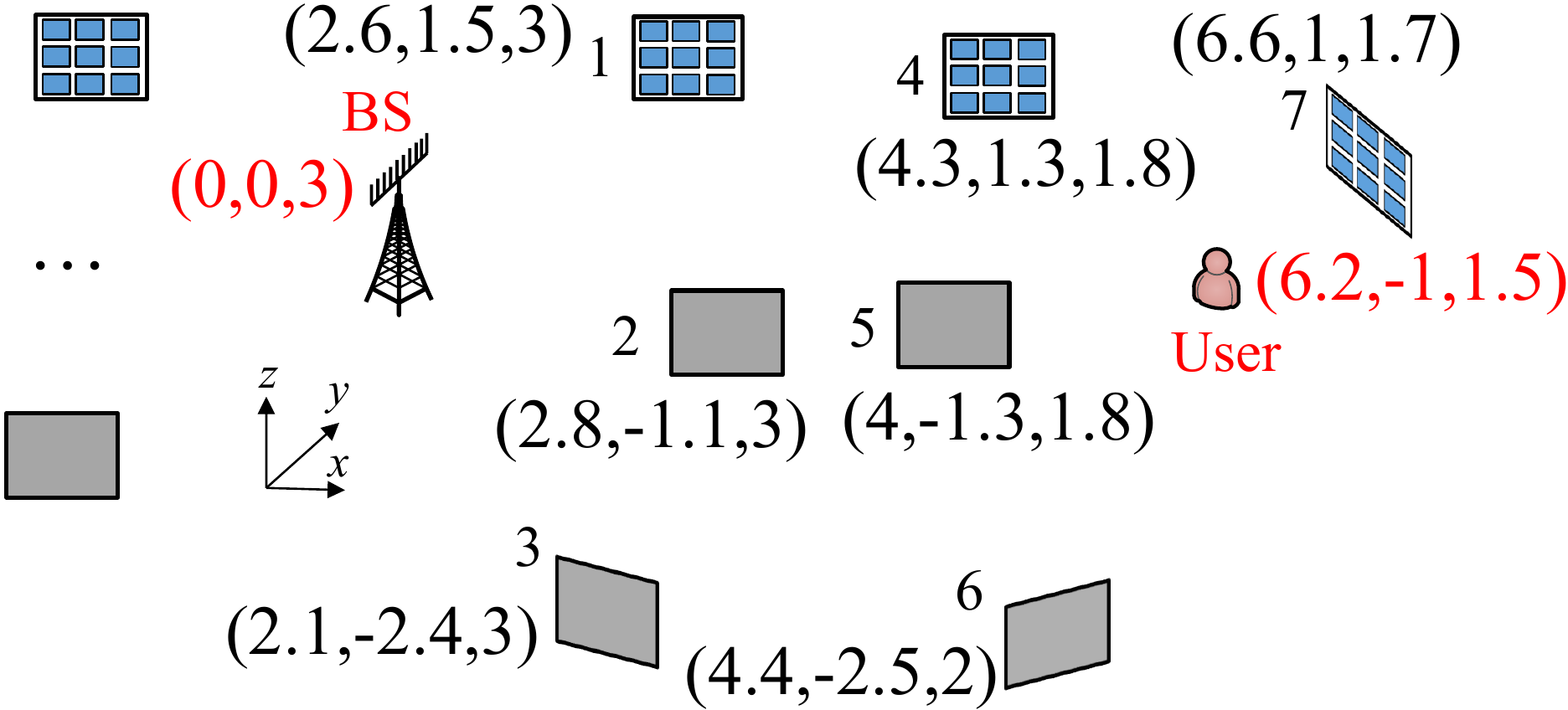}}
\subfigure[LoS graph.]{\includegraphics[width=0.24\textwidth]{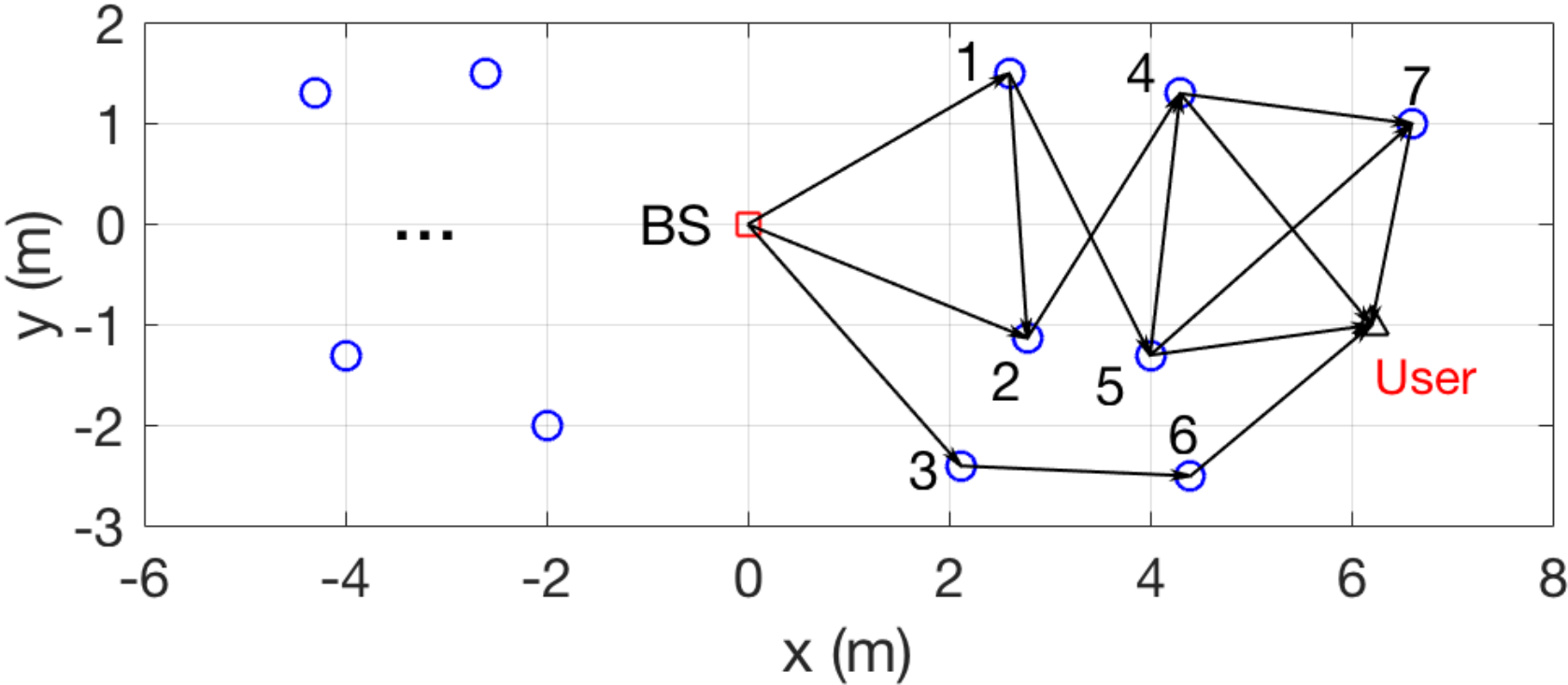}}
\vspace{-6pt}
\caption{Simulation setup of the multi-IRS system.}\label{topology}
\end{figure}
\begin{figure}[!t]
\centering
\subfigure[$M_0=18$]{\includegraphics[width=0.24\textwidth]{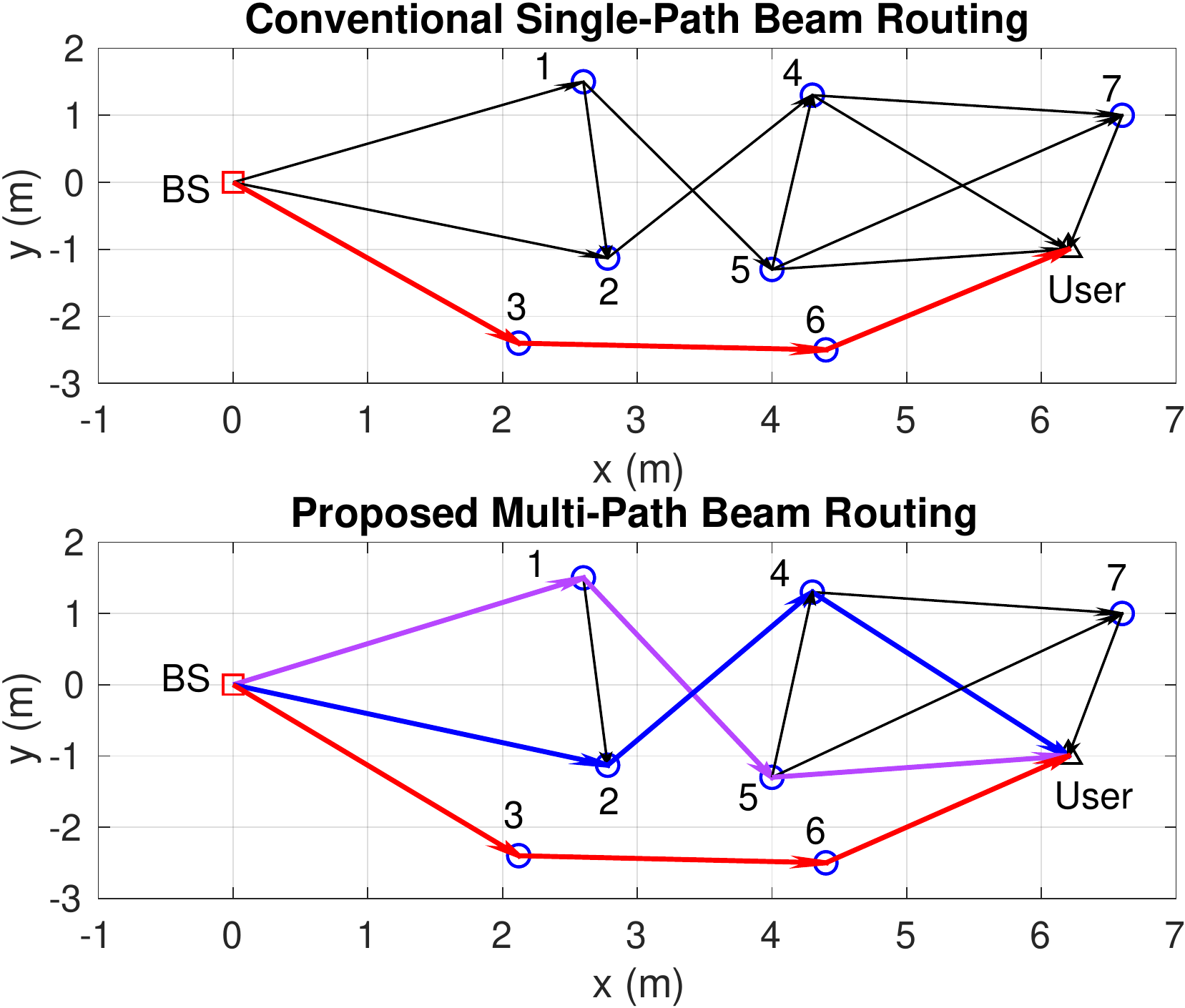}}
\subfigure[$M_0=24$]{\includegraphics[width=0.24\textwidth]{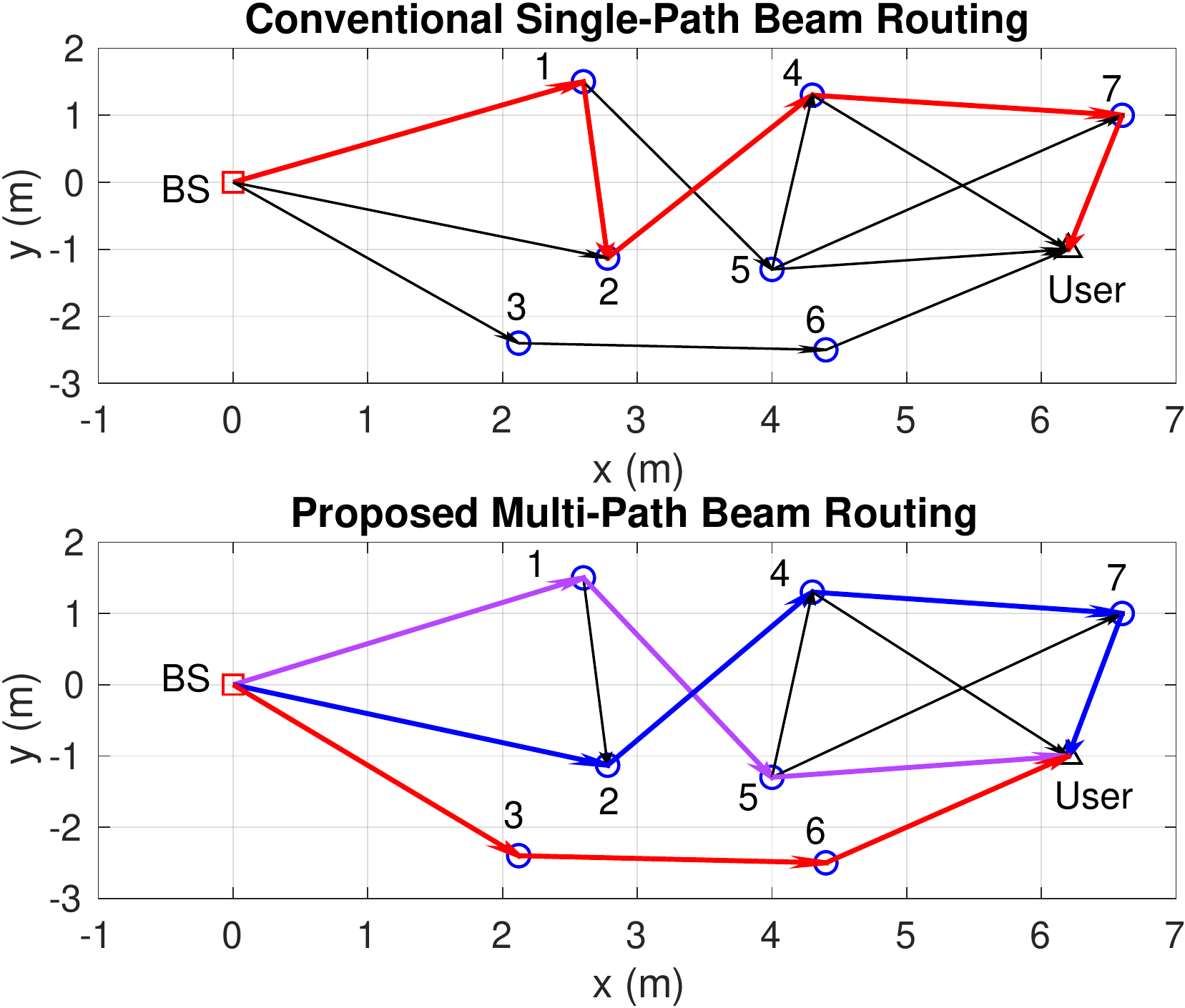}}
\vspace{-6pt}
\caption{Optimized reflection paths by single-/multi-path beam routing schemes with $K=4$.}\label{RefPath}
\vspace{-9pt}
\end{figure}
In this section, simulation results are provided to examine the performance of the proposed multi-path beam routing scheme in a complex indoor environment. The three-dimensional (3D) coordinates of all nodes involved and the facing directions of all IRSs in this system are shown in Fig.\,\ref{topology}(a), while the corresponding LoS graph $G_L$ is shown in Fig.\,\ref{topology}(b). Given the user location, it suffices to consider $J=7$ IRSs at the right-hand side of the BS in Fig.\,\ref{topology}. The number of BS antennas is $N_B=16$ with a spacing of half-wavelength. The carrier frequency is set to $f_c=5$ GHz and thus $\beta=-46$ dB. The numbers of elements in each IRS's horizontal and vertical dimensions are assumed to be identical as $M_0 \triangleq \sqrt{M}$, with a spacing of quarter-wavelength. Each IRS is assumed to apply 3D passive beamforming and use 64-point discrete Fourier transform (DFT)-based codebooks for its horizontal and vertical passive beamforming, respectively; while the BS uses 16-point DFT-codebook for its active beamforming. The transmit power of the BS is set to be 30 dBm.

\begin{figure}[hbtp]
\centering
\subfigure[]{\includegraphics[width=0.24\textwidth]{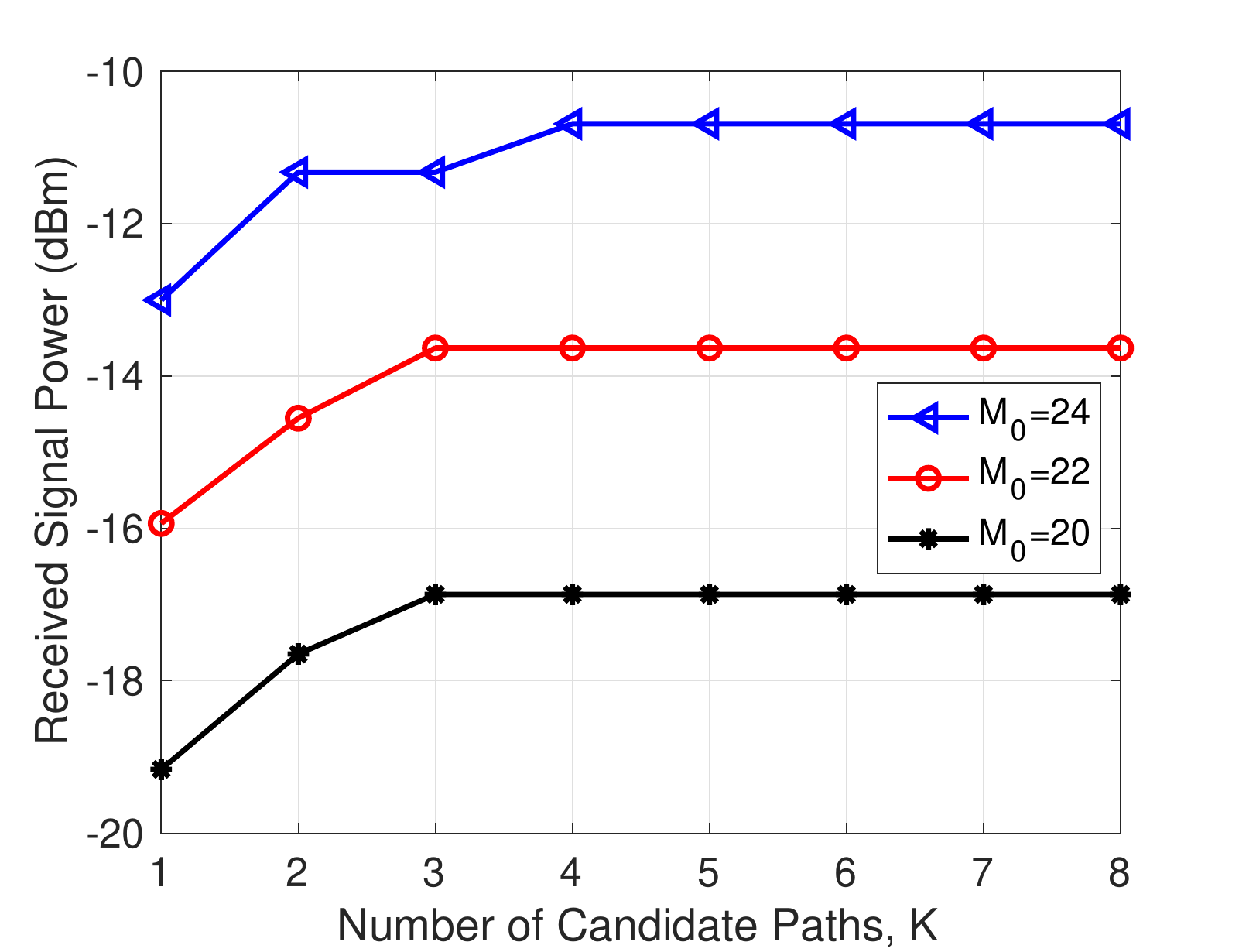}}
\subfigure[]{\includegraphics[width=0.24\textwidth]{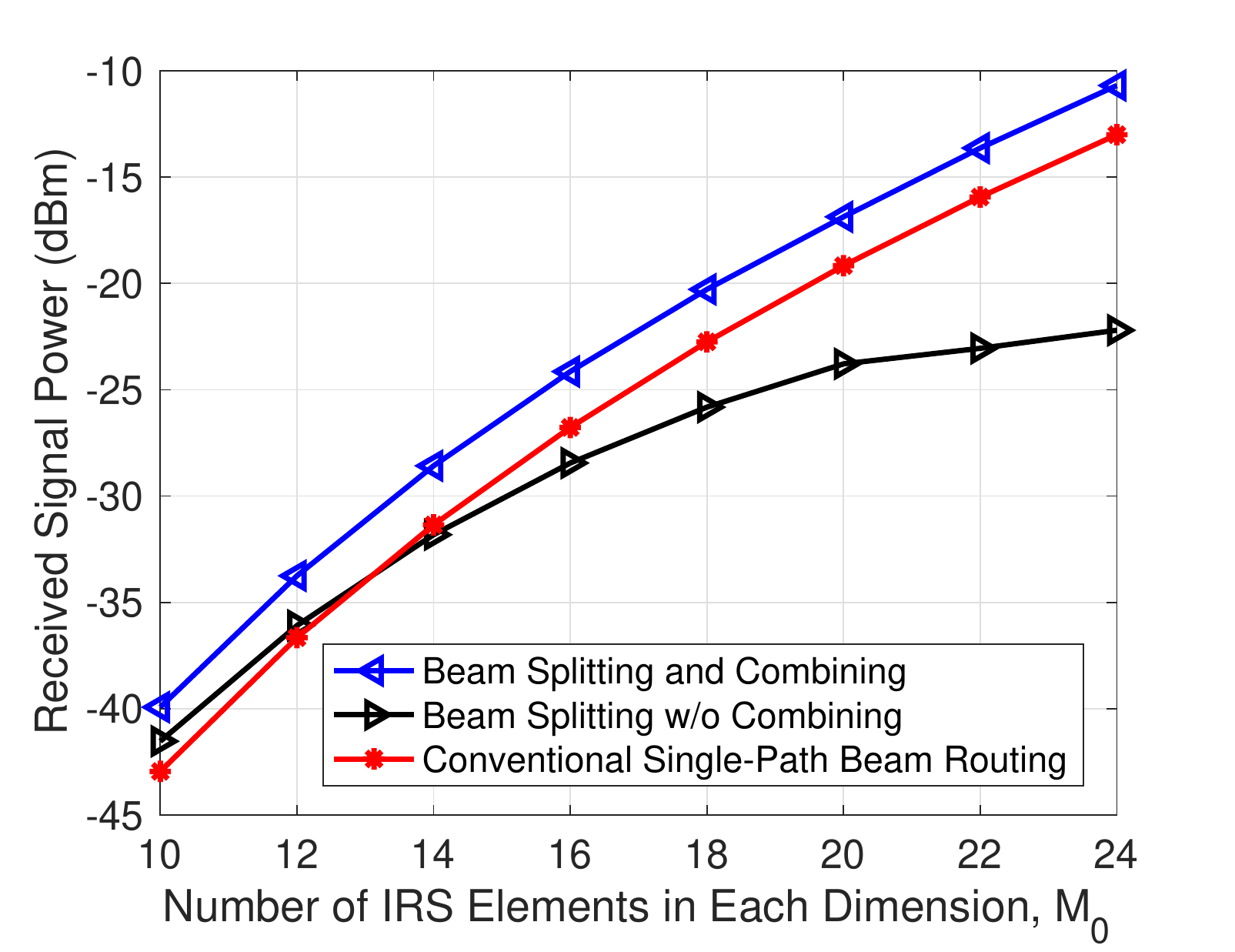}}
\subfigure[]{\includegraphics[width=0.24\textwidth]{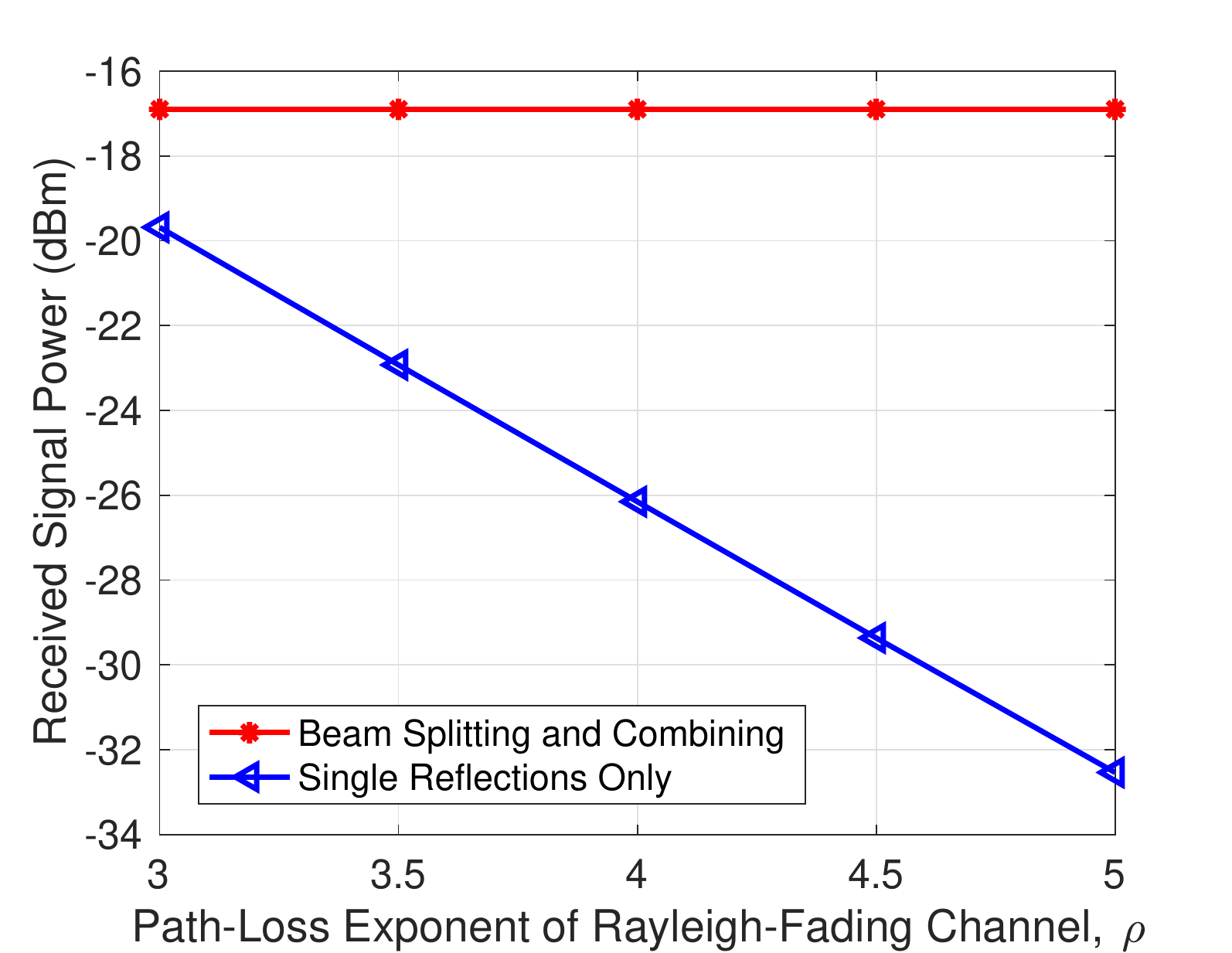}}
\subfigure[]{\includegraphics[width=0.24\textwidth]{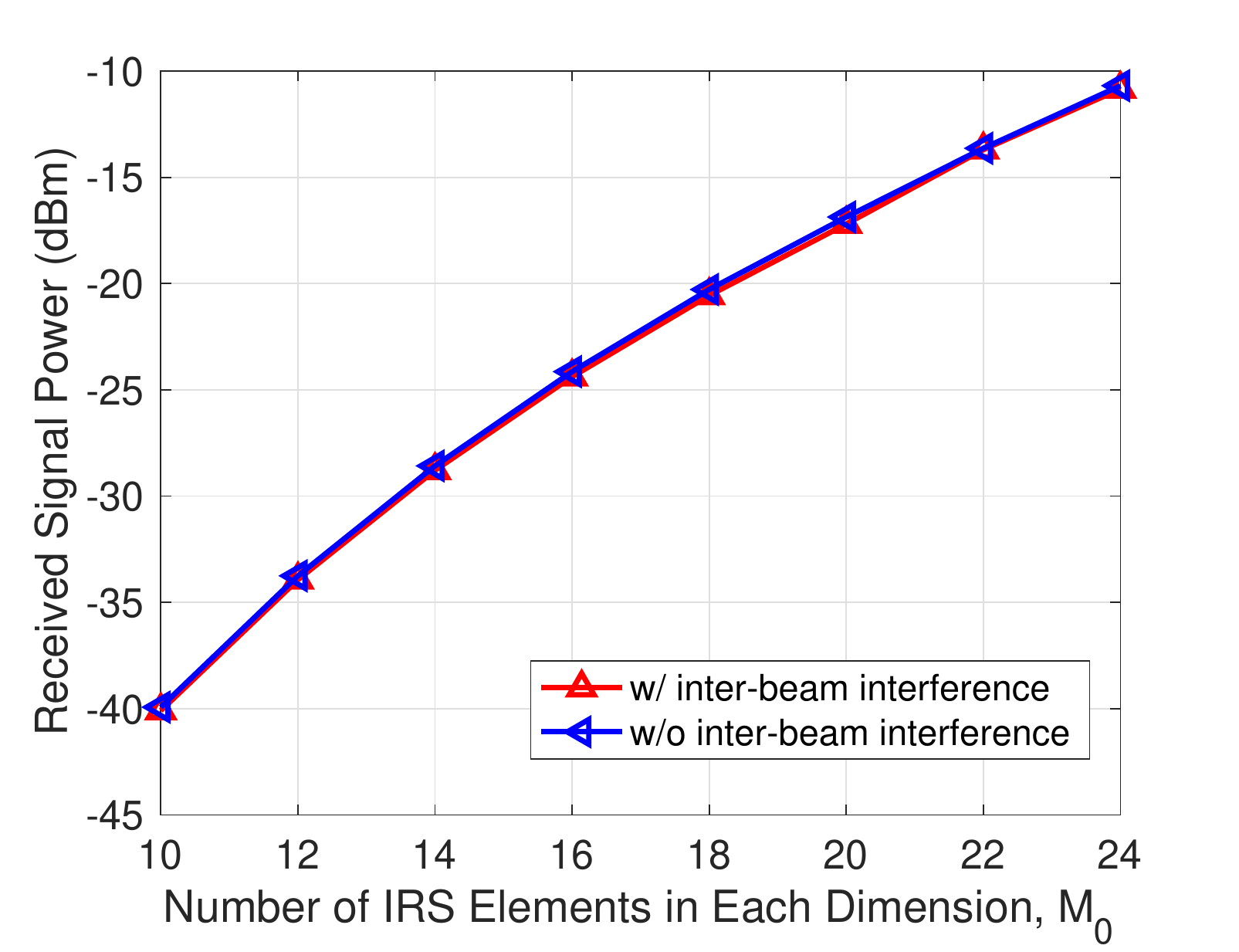}}
\vspace{-8pt}
\caption{Received signal power versus (a) number of candidate paths, $K$, (b) number of IRS elements in each dimension, $M_0$, and (c) path-loss exponent of Rayleigh-fading channels, $\rho$; (d) Received signal power with or without accounting for inter-beam interference.}\label{simfig}
\vspace{-12pt}
\end{figure}
First, in Fig.\,\ref{RefPath}, we plot the optimized reflection paths by the conventional single-path beam routing and the proposed multi-path beam routing. It is observed from Fig.\,\ref{RefPath}(a) that  under $M_0=18$, in the former case (a special case of the latter), the optimal reflection path only traverses two IRSs, i.e., IRSs 3 and 6, while leaving all other IRSs unused. In contrast, in the latter case, the BS splits the signal into $Q=3$ beams which are reflected to the user through three different paths, indicated by different colors. In particular, besides the path selected for the single-path beam routing (i.e., $\Omega_{\text{single}}^{\star}$), two additional paths are employed to improve the user's received signal power. However, $\Omega_{\text{single}}^{\star}$ is not necessarily selected for the multi-path beam routing, as shown in Fig.\,\ref{RefPath}(b) under $M_0=24$. 

Next, Fig.\,\ref{simfig}(a) shows the received signal power at the user receiver versus the number of candidate paths, $K$, in the proposed CBA under $M_0=18$. It is observed that the received signal power is monotonically non-decreasing with $K$, thanks to the enlarged solution set for (P1). It is also observed that the performance by CBA cannot be further improved by increasing $K$ when $K \ge 3$ under $M_0=20$ and $22$, and $K \ge 4$ under $M_0=24$. This implies that the optimal solution to (P1) is likely to be already found by CBA with a small $K$ in practice. Fig.\,\ref{simfig}(b) plots the received signal power at the user receiver versus $M_0$ by the single-/multi-path beam routing schemes. It is observed that the received signal power of the proposed multi-path beam routing scheme is approximately 3 dB higher than that of its single-path counterpart over the whole range of $M_0$. It is also interesting to note that if only the active beam splitting is performed but without the passive beam combining, i.e., the three beams in Fig.\,\ref{RefPath} are randomly combined at the user receiver, the resulting performance becomes even worse than the conventional single-path beam routing. 

In Fig.\,\ref{simfig}(c), we compare the proposed multi-path beam routing scheme with the conventional scheme which only exploits the single-reflection links from the BS to user via the $J$ IRSs under $M_0=20$. We assume Rayleigh fading for the NLoS links in the system, with a path-loss exponent of $\rho$. It is observed from Fig.\,\ref{simfig}(c) that thanks to the CPB gain, the proposed scheme yields a much better performance than the conventional one, especially when $\rho$ is large. Finally, to evaluate the effect of the inter-beam interference (e.g., over the LoS link from IRS 1 to IRS 2 in Fig.\,\ref{RefPath}(a)), we compare the user's received signal power at the user receiver with versus without accounting for the inter-beam interference in Fig.\,\ref{simfig}(d). It is observed that their gap is very small, which indicates that the effect of the inter-beam interference can be ignored.

\balance
\section{Conclusions}
In this letter, we propose a new multi-path beam routing scheme for multi-IRS aided wireless systems by exploiting the active beam splitting and passive beam combining techniques at the BS and user, respectively. As compared to the conventional single-path beam routing scheme, more available IRSs are utilized in the proposed scheme to further improve the user performance. We present the optimal BS/IRS active/passive beamforming design and BS power allocation for a given set of reflection paths and further propose a CBA to obtain a high-quality multi-path beam routing solution. It is shown via simulation that the proposed scheme can yield significant performance gains over its single-path special case.

\bibliography{BeamSplit.bib}
\bibliographystyle{IEEEtran}
\end{document}